# Glass softening, crystallization, and vaporization of nano-aggregates of Amorphous Solid Water: Fast Scanning Calorimetry studies

Deepanjan Bhattacharya, Liam O'Reilly, and Vlad Sadtchenko[*]

Department of Chemistry

The George Washington University

Washington, DC 20052

[*]Corresponding author; Chemistry Department, The George Washington University, 725 21st St. NW, Rm. 107, Washington, DC 20052; phone: 202-994-6155; e-mail: vlad@gwu.edu

**ABSTRACT**

Fast scanning calorimetry (FSC) was employed to investigate glass softening dynamics in amorphous solid water (ASW) nano-aggregates with thicknesses ranging from 2 to 20 nm. ASW nano-aggregates were prepared by vapor-deposition on the surface of a tungsten filament near 141 K and then heated at a rate of $1.1 \cdot 10^5$ K/s. The resulting thermograms' complex endo- and exothermal features were analyzed using a simple model. The results of the analysis show that glass softening of ASW nano-aggregates takes place at 160 K and vaporization of ASW nano-aggregates can take place at temperatures as low as 185 K. The results of these studies are discussed in conjunction with results of past studies of glass softening dynamics in water in various confining geometries.

**I. INTRODUCTION**

Water plays central role in myriad of biological and environmental processes [1-3]. Under a variety of conditions, the most important and interesting phenomena occur not in bulk of water samples, but at liquid-vapor and liquid-solid interfaces [4]. Furthermore, in many important physical, chemical, biochemical, and environmental processes, water is present in the form of sub-nanometer clusters, nanoscale solid or liquid particles, and ultrathin films, *i.e.*, under condition of extreme confinement[5]. Examples of such processes include protein and nucleic acid hydration[6], environmentally important chemical reaction in quasi-liquid layers in surfaces of stratospheric polar clouds[7]; and interstellar and interplanetary chemistry where water may be present in highly microporous amorphous form[8].

Due to its role in broad range of natural phenomena, confined water has been the focus of intense experimental and theoretical scrutiny for decades. In addition to applied interest in understanding properties of water under confinement is important for developing a comprehensive fundamental picture of relationships between various condensed phases of $H_2O$. Microscopic $H_2O$ aggregates are uniquely amendable to modern computational studies. In fact, ASW nanoparticles and smaller clusters are often considered as an idealized simplified systems for theoretical investigations [8-14] which results can potentially be used to understand properties of bulk-like hydrogen bonded systems.

Despite significant experimental and theoretical work carried out over past decades, understanding structure and dynamics in various confined and bulk-like condensed phases of water has posed a scientific challenge [15]. For instance, the assignment of glass transition temperature ($T_g$) of water to 136 K has been intensely debated [1, 3, 16]. Experimental studies in support of the assignment of glass transition of water at 136 K are numerous and include traditional Differential Scanning Calorimetry (DSC) experiments with pure water and aqueous solutions[17, 18], Broadband Dielectric Spectroscopy (BDS) measurements of relaxation times in water saturated Poly(HEMA) matrix[19], investigations of water's mechanical properties[18] at temperatures near 140 K, Vibrational Spectroscopy studies of supercooled aqueous solutions[20], Temperature Program Desorption (TPD) experiments with ultrathin (< 30 nm) films of vapor-deposited water[21], and Time-of-Flight Secondary Ion Mass Spectrometry (TOF-SIMS) investigations of nanoscale water films on various substrates[22-25]. These and other experimental results support the notion that water is essentially a viscous liquid (i.e. is characterized by structural relaxation times shorter than 100 s) in the temperature range of 10 to 20 K *prior* to the onset of rapid crystallization.

The alternative view of water at temperatures below 228 K emphasizes its possible glass-like (solid) state which is characterized by kinetic and thermodynamic parameters similar to those of a crystalline ice. For example, Angell and coworkers argued consistently that postulated $T_g$ of water is incorrect, and must be reassigned to a higher value.[1, 16, 26, 27] It was also suggested that pure, crystallization of bulk-like pure water samples takes place before they can undergo a glass softening transition. Using fast scanning calorimetry, Bhattacharya et al. showed that although nanoscale films of ASW undergo an observable softening transition prior to crystallization, the bulk-like pure ASW samples do not. Chonde *et. al.* and Mc Cartney *et. al.*, demonstrated that molecular impurities may result in significant decrease in $T_g$ of doped glassy water samples, as compared to that of pure water [28]. Dielectric spectroscopy studies of Minoguchi *et. al.* seem to deny the existence of viscous liquid water[29]. Mullins and coworkers studies showed that the abnormally high (liquid-like) values of water self-diffusivity measured by Smith *et. al.,* at temperatures near those characteristic of the onset of crystallization may be explained by rupturing of the ultrathin glassy water films in the course of concurrent crystallization [30, 31]. This conclusion is in agreement with the latest findings of the Oguni's group [32]. Jankowiak and coworkers[33] conducted single molecular spectroscopy experiments with hyperquenched glassy water, which indicate that water is *not* a viscous liquid at temperatures above 140 K. These results are in contrast to the conclusions of the ESR experiments [34] where a liquid-like dynamics of a probe species were observed at similar temperatures. Calorimetric studies of Oguni *et. al.* suggests that $T_g$ of bulk water may be as high as 210 K [35].

As we argued in the past, a set of contrasting experimental results may be explained by an assumption that an important experimental parameter may be unaccounted for in all these studies giving rise to conflicting findings [36-38]. In particular, we argued that structural "defects" in the bulk of glassy water samples, such as micropores or fissures, may serve as dynamic "hot spots" where $H_2O$ mobility is greatly enhanced. According to our hypothesis, supported by our experimental findings and by theoretical investigations, water in the vicinity of such dynamic heterogeneities is capable of undergoing a glass softening transition at temperatures tens of degrees below the characteristic temperature of the onset of crystallization.

In the past, we reported results of Fast Scanning Calorimetry (FSC) investigations of $H_2O$ molecular dynamics at interfaces between ASW and other "soft" solids. Taking into account, fundamental and applied significance of confined water, a study of free-standing ASW nanoaggregates (with effective thicknesses as low as 2 nm) was undertaken as a logical continuation of experiments described in Ref. [36]. The goals of these studies were two-fold: *first,* to validate finding of FSC experiments with stratified ASW film; *second*; to explore kinetics and thermodynamics in nano-confined water by means of heat capacity spectroscopy (a general term for DSC, FSC or AC calorimetry techniques).

## I. EXPERIMENTAL

### A. Overview.

Experiments were conducted using custom built Fast Scanning Calorimetry (FSC) apparatus. Detailed description of this setup is given elsewhere [36-41]. Therefore, only a brief account will be given here. As shown in Fig. 1, the central part of the microcalorimeter setup is a bare tungsten filament which is 1.5 cm long and 10 μm in diameter. The filament is soldered to two supports, which are in thermal contact with liquid nitrogen (LN)-cooled heat sink through 1 mm thick sapphire plates. While providing a moderate thermal contact with the heat sink, the sapphire plates ensured that the filament and the supports are electrically isolated from the rest of the apparatus. Two resistive heaters capable of dissipating up to 10 W of power were attached to the filament supports. The temperature of the filament can be varied from 100 to 170 K by balancing of the heat load from the resistive heaters against thermal conductance to the LN-cooled heat sink. The temperature of the filament supports can be monitored with two miniature thermocouples (T type). The microcalorimeter assembly with the effusive dosers is surrounded by six cryogenic shields and positioned inside a vacuum chamber pumped equipped with a 1500 L·s$^{-1}$ turbomolecular pump. During all experiments, the chamber was maintained at the base pressure of 5·10$^{-7}$ Torr. The pressure inside the chamber was monitored using an ionization gauge.

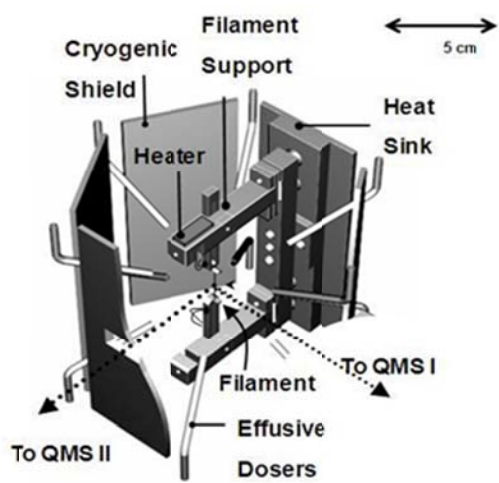

FIG.1. Schematic of the Fast Scanning Calorimetry-Fast Thermal Desorption Spectroscopy (FSC-FTDS) apparatus: Filament supports, cryogenic shields, and vapor deposition dosers.

As shown in Figure 1, the vapor deposition system for the preparation of ASW films consists of 12 effusive dosers. The dosers are 1/8″ in diameter, which are equally spaced around the filament at the distance of about 2 cm from it. Each tube is positioned at the angle of about 60° with respect to the filament axis. The dosers are connected to the external vapor source, which consists of a vial filled with HPLC grade water and maintained at 291 K. The vial is also connected to a mechanical pump through LN trap. The pumping speed of the mechanical pump can be adjusted with a needle valve. Such an arrangement makes it possible to avoid pressure build up inside the vial when the vapor source is not in use. The vapor deposition time can be varied from 50 ms to several hours.

A typical FSC experiment started with the deposition of nanoscale film of ASW on bare tungsten filament at temperatures near 141 K. This deposition temperature ensured low porosity and crystallinity of the ASW films. During deposition, the filament temperature was maintained within ± 1 K of the set point (141K). After the film deposition, it was annealed for 4 s at 141K, and then heated at a rate of $1.1 \cdot 10^5$ K/s by applying 2.6 V potential difference across the filament. The final temperature of the film-filament system typically reached 330 K. The maximum temperature of 330 K ensured that the film completely vaporized at the end of each FSC scan. After the rapid heating, the filament was allowed to cool down to 141 K, which took less than two seconds.

During rapid heating, the current and voltage data were recorded every 2 μs. The current and voltage data were used to infer the power generated by the filament and its temperature-dependent resistance. The temperature of the filament was calculated from the resistance data using a calibration procedure described later in this section. The apparent heat capacity of the ASW covered filament was calculated as the ratio of the power generated to the first time derivative of temperature. The heat capacity data was then plotted against temperature, resulting in thermograms similar to those measured during traditional differential scanning calorimetry. In order to obtain thermogram of ASW layer only, thermogram of bare filament was measured and subtracted from the thermogram of ASW covered filament.

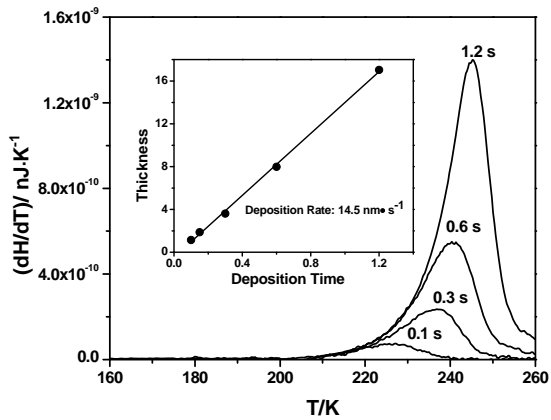

**FIG. 2.** Thermograms of ASW films vapor deposited on tungsten films at temperatures near 141 K. The large endotherms are consistent with the onset of film vaporization, as verified by MS. Analysis of the endotherms gives an estimate of the vaporization enthalpy and thickness of the films. The inset shows the plot of thickness of the film vs. deposition time.

In order to ensure accurate temperature measurements during FSC experiments, a special calibration procedure was implemented at the beginning of each day. The filament assembly was slowly cooled down from ambient to cryogenic temperatures. In the course of cooling, the temperature of the filament assembly was measured by the two thermocouples attached to filament supports. The temperature of the filament was assumed to be equal to the temperature of the filament supports during slow cooling. At the same time, the resistance of the filament was measured with an ohm-meter. The temperature of the filament was plotted as a function of its resistance, and was fitted with a straight line. The slope and intercept of the fit line was used to calculate temperature values from the resistance values obtained during a FSC scan. Systematic errors in temperature measured were less than 2 K for daily experiments.

**B. Film thickness determination.**

The effective thickness of the ASW aggregates was derived from the heat of vaporization. Fig. 2 shows the thermograms of ASW films vapor deposited on tungsten filament at temperatures near 141 K. The large endotherms, at temperatures above 200 K, are consistent with the onset of $H_2O$ vaporization from the filament surface, as verified by mass spectroscopic analysis (Fig. 4). As shown in Fig. 2, an increase in the deposition time results in a proportional increase in the magnitude of the endotherm. Taking into account the fact that ASW films typically undergo rapid crystallization at temperatures above 200 K, we conclude that the endotherms in Figure.3 are due to vaporization of crystalline ice films.

The vaporization endotherms shown in Fig.2 were used to determine the effective thickness of nanoscale films with high accuracy. First, the net enthalpy of vaporization of an ASW film was estimated by numerically integrating the vaporization endotherm in the temperature range from 200 to 260 K. Second, the mass of the film was calculated by dividing the net enthalpy of vaporization by specific vaporization enthalpy of crystalline ice (52 KJ·mol$^{-1}$). Finally, the thickness of each ASW film was estimated from its mass by assuming that its density was near 0.98 g·cm$^{-3}$ and that the film had a neat cylindrical geometry. The results of this analysis are illustrated in the inset of Fig. 3 that shows film's thickness as a function of $H_2O$ vapor- deposition time. From the slope of this plot the deposition rate was estimated to be 15± 1 nm·s$^{-1}$. The linear nature of the plot shows that the vapor source gave a stable deposition rate.

Heating rates in excess of hundred thousand kelvins per second employed during FSC experiments may raise concerns about the influence of possible thermal lags and gradients on the results of our experiments. However, we emphasize that, despite high heating rates, the temperature lags and gradients must be less than a fraction of a degree. We have already addressed the issue of thermal lags and gradients in micrometer scale films of ASW, in our previous publications,[36-38] and found it to be less than a few degrees. Also note, in the present study, experiments are conducted with *nanoscale* films, which should result in negligible thermal lags and gradients compared to micrometer scale films. The thermal equilibration time of a nanoscale film is negligible compared to typical time scale of FSC experiments. In short, thermal lags and gradients for FSC studies of nanoscale films of ASW can be neglected.

## III. RESULTS AND DISCUSSION

### A. Thermograms of 2 nm thick films of ASW nanoaggregates.

Fig. 3 shows FSC thermogram of 2±1 nm nano-aggregates of ASW. The deposition temperature was chosen near 141 K in order to ensure that the ASW film was well annealed while not undergoing extensive crystallization during vapor deposition. The deposition time was 0.15 s, annealing time was 4 s and the heating rate was approximately $10^5$ K/s. The thermogram in Fig. 3 consists of three endotherms and two exotherms. Endotherm 1, with an onset at 210 K, and endotherm 3, with an onset at 185 K, and exotherm 2 with an onset at 200 K were observed. As discussed in the experimental section, the large endotherm at temperatures above 200 K is consistent with the onset of $H_2O$ vaporization from the filament surface. The inset of Figure 3 shows a high signal-to-noise ratio FSC thermogram of the ASW film in the temperature range from 140 to 210 K. The high signal-to-noise ratio was achieved by the averaging of 2000 individual thermograms of nanoaggregates of ASW. Two endotherms and an exotherm are visible. Endotherm 3, with an onset at 185 K, endotherm 5 with an onset at 160 K and exotherm 4 with an onset at 175 K. Heat capacity of water (dashed line)[32] and ice (dotted line)[42] are shown for comparison with that of ASW.

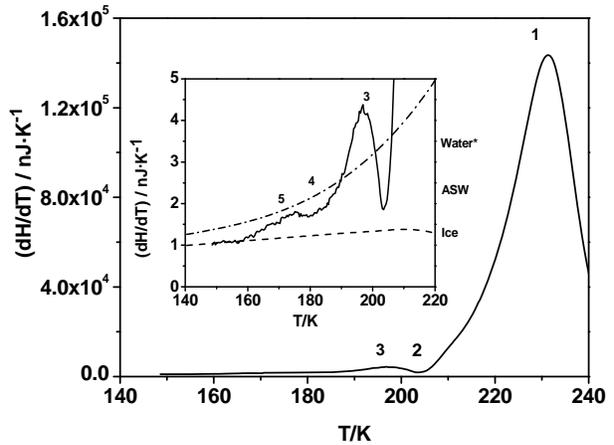

**Figure. 3.** FSC thermogram of 2±1 nm nano-aggregates of ASW. Inset: High S/N ratio FSC thermogram (average over 2000 individual thermograms) of 2±1 nm nano-aggregates of ASW. The deposition temperature was chosen near 141 K in order to ensure that ASW film was well annealed and also without undergoing extensive crystallization. The deposition time was 0.15 s, annealing time was 4 s and the heating rate was around $10^5$ K/s.

**B. The model.**

In order to disentangle the contribution of various endo- and exothermal processes to the complex thermograms of nanoaggregates of ASW, we developed a simple model which takes into account not only possible glass softening transitions but also vaporization of the ASW samples. We begin first with the contribution of vaporization of water, $\left(\frac{dH}{dT}\right)_{vap}(T)$, to the overall thermogram. During the derivation of an expression for $\left(\frac{dH}{dT}\right)_{vap}(T)$, it was assumed that the filament has a geometrical surface area, $S_{fil}$, which can dissipate heat by vaporization of ASW nanoaggregates. Nevertheless, the nanoaggregates of ASW may have complex morphology resulting in differences between geometrical area of neat cylinder and the effective surface area of the ASW film. The likely difference between the surface area of the filament and that of the ASW film was taken into account by introducing the effective surface area of nanoaggregates, which is $\alpha \cdot S_{fil}$, where α is a coefficient. If the mass accommodation coefficient of water on ASW film is unity, then the vaporization rate, $R_{vap}(T)$, of the entire ASW film can be expressed as

$$R_{vap}(T) = \alpha \cdot S_{fil} \cdot \frac{P_{H_2O}(T)}{\sqrt{2\pi \cdot m \cdot kT}} \,, \tag{1}$$

where $P_{H_2O}(T)$ is the equilibrium vapor pressure of either ASW or crystalline ice. Note that the vapor pressure of ASW is approximately *twice* of that of crystalline ice at temperatures ranging from 130 to 150 K.[43]

The heat flow from the filament to ASW which leads to ASW's vaporization can be estimated using the following expression:

$$\left(\frac{dH}{dt}\right)_{vap}(T) = R_{vap}(T) \cdot \Delta H_{vap}(T), \qquad (2)$$

where $\Delta H'_{vap}(T)$ is the specific (per molecule) enthalpy of vaporization of either ice or amorphous solid water[43]. Thus, the vaporization endotherm, $\left(\frac{dH}{dT}\right)_{vap}(T)$, can be simulated as

$$\left(\frac{dH}{dT}\right)_{vap}(T) = \left(\frac{\partial H}{\partial t}\right)_{vap} \cdot \left(\frac{dt}{dT}\right), \qquad (3)$$

where $\left(\frac{dt}{dT}\right)$ represents the inverse of the heating rate.

Combining equations 1 and, we obtain the following expression for the vaporization endotherm,

$$\left(\frac{dH}{dT}\right)_{vap}(T) = \alpha \cdot S_{fil} \cdot \frac{\Delta H_{vap}(T) \cdot P_{H_2O}(T)}{\sqrt{2\pi \cdot m \cdot kT}} \cdot \left(\frac{dt}{dT}\right). \qquad (4)$$

Note that $\alpha$ is essentially the only fitting parameter used in the equation 4.

In the entire temperature range of the thermograms (from 140 K to 250 K), vaporization of ASW and crystalline ice (at lower and higher temperatures, respectively) was modeled as a combination of amorphous ice and crystalline ice vaporization data,

$$\left(\frac{dH}{dT}\right)_{vap} = \left(\frac{dH}{dT}\right)_{Vap}^{ASW} \cdot S^+ + \left(\frac{dH}{dT}\right)_{Vap}^{ice} \cdot S^-, \qquad (5)$$

where $\left(\frac{dH}{dT}\right)_{Vap}^{ASW}$ and $\left(\frac{dH}{dT}\right)_{Vap}^{ice}$ are the vaporization endotherm due to amorphous solid water and crystalline ice, respectively and $S_i$ is the sigmoidal function. The sigmoidal function is of the form,

$$S_i^{\pm} = 1 + e^{\pm \Delta_{Cr}(T - T_{Cr})}, \tag{6}$$

where $T_{Cr}$ is the temperature minimum of the crystallization exotherm (205 K) and the $\Delta_{Cr}$ is a parameter which describes approximately the width of the exotherm.

Having taken into account the contribution of vaporization of water to the overall thermogram, we now proceed to model the heat capacity contribution to the overall thermogram. The heat capacity of nanoaggregates of ASW, in the temperature range of 140- 230 K, was modeled as a combination of ice and liquid water heat capacities,

$$C_{film} = C_{ice} \cdot S_i^+ + (\chi_{SF} \cdot C_{ice} + \chi_{LF} \cdot C_{LF}) \cdot S_i^- , \tag{7}$$

where $C_{ice}$ is heat capacity of ice, $C_{liq}$ is heat capacity of liquid water, $\chi_{SF}$ is the fraction solid fraction of the film, and $\chi_{LF}$ is the liquid fraction of the film. The heat capacity of liquid water was that of water confined in nanopores.[32] The heat capacity of ice was obtained from literature values.[42] $S_i$ is the sigmoidal function. The sigmoidal function was used to simulate the changes in the heat capacity during possible glass softening of ASW, and is described by the following expression,

$$S_i^{\pm} = 1 + e^{\pm \Delta_{tr}(T - T_{Tr})}, \tag{8}$$

where $T_{Tr}$ is the temperature of transition and $\Delta_{tr}$ is the width of a transition from amorphous solid film to *partially* softened ("melted") film during FSC scan.

**C Analysis of the ASW thermograms.**

Despite including many components, the model described in the previous section can be used successfully to disentangle various endo- and exothermal processes during rapid heating of ASW nanoaggregates. The simplicity of applying the model for analysis of the ASW thermograms is due to great separation of distinct thermal phenomena on the temperature scale in the FSC thermograms. As shown in figure 4 (double dotted-dashed line), the heat loss due to vaporization is the dominant contribution to the FSC thermogram at temperatures above 210 K. At this temperatures, variations in the actual heat capacity of the film (e.g., due to mass loss by vaporization) can be safely neglected. At the same time, contribution of film's vaporization to overall effective heat capacity is insignificant at temperatures below 180 K. Furthermore, at temperatures below 220 K, the fraction of the film which has vaporized is less than a few percent, i.e., the vaporization has no impact on actual heat capacity of the film. In short, the extrema of the temperature range of the ASW nanoaggregates thermograms can be adjusted to the distinct components of the system with ease and high accuracy. The accurate fits of the thermogram data in the limit of low and high temperature can then be extrapolated to the intermediate temperature range where the interplay of distinct endo- and exothermal contributions matters.

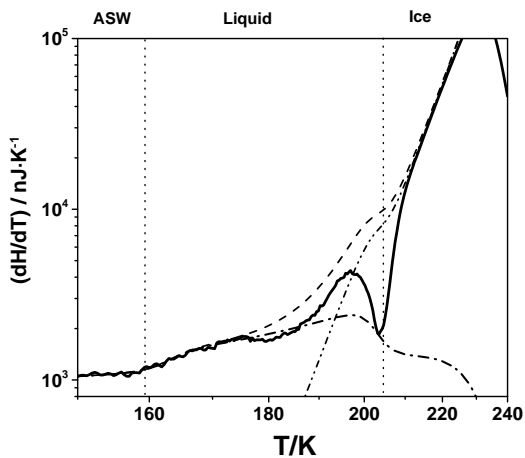

**Figure. 4.** High S/N ratio FSC thermogram (average over 2000 individual thermograms) of 2±1 nm nano-aggregates of ASW (Black line). The deposition temperature was chosen near 141 K in order to ensure that ASW film was well annealed and also without undergoing extensive crystallization. The deposition time was 0.15 s, annealing time was 4 s and the heating rate was around $10^5$ K/s. Black single dotted-dashed line represents the modeled heat capacity of water. Single dash- double dotted line depicts the vaporization endotherm of water. Black dashed line represents combination of modeled heat capacity and vaporization endotherm data.

Figure 4 compares the experimental heat capacity of the nanoaggregates of water (solid line) with that of the model (dashed line). As shown in figure 4, in the temperature range from 140 to 160 K, the heat capacity of nanoaggregates of ASW neatly follows that of heat capacity of ice. A gradual increase in the heat capacity of the nanoaggregates of ASW is observed at temperatures above 160 K, reaching a maximum near 175 K. The magnitude in the rise in the heat capacity is consistent with literature value of liquid water confined in nanopores.[32] under assumption that approximately 70 % of the 2 nm thick ASW sample undergoes softening. In other words, according to the analysis, significant but not complete transformation of 2 nm thick ASW film takes place at temperatures above 160 K.

At temperatures near 175 K, a weak exotherm is observed. The possible origin of this weak exotherm may be due to Ostwald ripening of clusters of nanoaggregates of water[9]. The weak exotherm is followed by an endotherm at temperatures from 185 K to 195 K. Note that the vaporization rate of ASW in the temperature range from 185 to 205 K becomes significant and is twice of that of crystalline ice in the same temperature range. Also note that the shape of the observed endotherm is very similar to the shape of the model heat capacity (dashed line) which is dominated by vaporization term in this temperature range. In short, the endotherm near 180 K is most likely to originate from vaporization of ASW. This conclusion is further validated by analysis summarized in Fig. 4.

Fig. 4 shows the *adjusted heat capacity* of 2 nm thick ASW, i.e., it shows the result of subtraction of model vaporization endotherms from FSC thermogram data. As shown in the figure, no residual endotherms are evident in the plot of the excess heat capacity at temperatures above 170 K.

**D. Surface area of ASW nanoaggregates of distinct thicknesses.**

Before we can proceed with analysis of FSC thermograms of ASW nanoscale aggregates, we must obtain an estimate of relative surface areas of films of different mass and effective thicknesses. Taking into account the high surface roughness of tungsten filament and its likely hydrophobicity (tungsten surfaces under high vacuum conditions tend to form graphite films), the nanoscale ASW ad layers are highly unlikely to have neat thin film geometry and may consists of nanoscale clusters. The increase in mass of such ASW deposits must be accompanied by the increase in the average cluster size, and therefore in the overall surface area of the deposit. As we explain in the next section of this article, the variation in the surface area with average thickness of ASW deposits is an important parameter for determination of microscopic location where ASW softening occurs (e.g., surfaces of clusters vs. bulk).

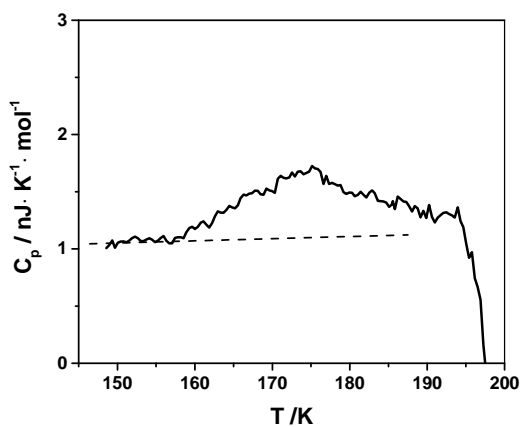

**Figure. 4.** The *adjusted heat capacity* of 2 nm thick ASW. The thermogram was obtained by subtracting model vaporization endotherms from FSC thermogram data. As shown in the figure, no residual endotherms are evident in the plot of the excess heat capacity at temperatures above 170 K.

As illustrated by Fig. 4, due to high sensitivity of FSC towards endothermic process of ASW vaporization, the vaporization of the exceedingly thin (less than a few percent) top layer of ASW nanoaggregates can be detected. The heat lost during the onset of vaporization must be proportional to overall surface area of the film, and can be used for estimates of variation of the surface area with the mass or effective thickness of the deposit.

Fig. 5 shows the high temperature part of FSC thermograms of nanoaggregates of ASW films, which thicknesses range from 2 to 20 nm. The high temperature part of the FSC thermogram consists mainly of crystallization exotherm and vaporization endotherm. The vaporization endotherm at temperatures consistent with complete crystallization of the sample was fitted with equation 4, which in this temperature range, represents heat loss due to vaporization of crystalline ice only. However, the extrapolation of the model towards temperatures below crystallization exotherm makes it possible to determine contribution of ASW vaporization to the overall thermogram.

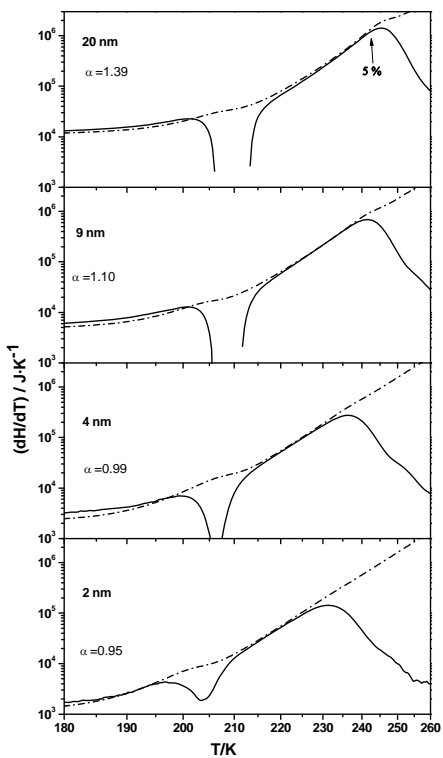

**Figure. 5.** High temperature FSC thermograms of films of thicknesses 2, 4, 9 and 20 nm, respectively. Each of the thermograms is fitted with crystalline ice vaporization data. Note, α represents the effective surface area

As the model for heat loss due to vaporization of crystalline ice neatly fits the vaporization endotherm of nano-aggregates of ASW, in the temperature range of 210-240 K, it can be concluded that zero-order vaporization kinetics is observed in this temperature range. As already noted, the effective surface area, α, is the only fitting parameter in equation in this analysis, and represents the relative surface area of the nanoaggregates of ASW to the geometrical surface area of the filament. By varying the magnitude of the fitting parameter α for films ranging in thicknesses from 2 nm to 20 nm, the surface area of the nano-aggregates of ASW relative to the geometrical surface area of the filament can be estimated.

Figure 6 summarizes the analysis described immediately above and shows a plot of film's relative surface area as a function of *effective* thickness of the film. The relative surface area dependence on effective film thickness can be approximated by a linear function: the increase in the effective thickness from 2 to 20 nm results in an increase in the surface area of the film by approximately 50%.

**E. Glass softening of ASW nanoaggregates: surface vs bulk**

As we have already discussed, according to preliminary analysis, only 70 % of the film undergoes glass softening at temperatures above 160 K in 2 nm thick nano-aggregates of ASW. The glass softening may occur either at the surface or bulk of the film. In order to understand where the glass softening may be occurring, i.e., in order to determine whether or not the ASW glass softening is a purely surface phenomenon, an analysis of thermograms of films of ASW, with thickness ranging from 2 to 20 nm, was carried out. Under certain circumstances, the analysis of results from experiments with films of varying thicknesses is a straightforward approach to distinguish between surface and bulk phenomena. Indeed, if a process originates at the surface of neat thin film, the observables must be independent on the film thickness. In the case of the ASW nanoaggregates, however, the overall surface area increases along with overall volume of the deposit, albeit not as rapidly. Thus a special analytical procedure was designed to take into account the variations in the surface area of ASW samples of different effective thicknesses.

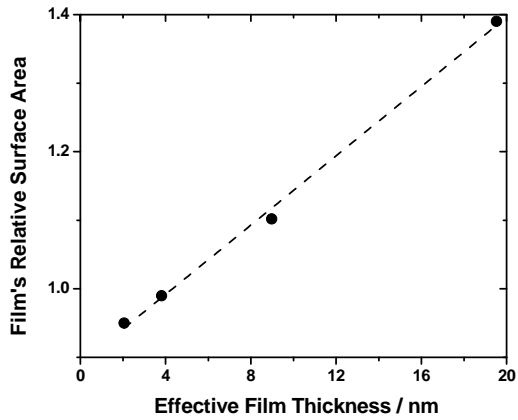

**Figure. 6.** Variations in the overall film's relative surface area with effective film thicknesses. See text for details.

First, the heat capacity of ice and the heat loss due to vaporization were subtracted from the overall thermogram of nano-aggregates of ASW. The result of the subtraction is *the excess heat capacity* of ASW aggregates.

$$\Delta C_p^{film} = C_p^{film} - M_{film} \cdot C_p^{ice*} - \left(\frac{dH}{dT}\right)_{vap}, \qquad (9),$$

where $C_p^{film}$ is the measured heat capacity of the film, $C_p^{ice*}$ is the specific heat capacity of ice, $M_{film}$ is the mass of the film, and $\left(\frac{dH}{dT}\right)_{vap}$ is the heat loss due to vaporization.

Second, the experimental values of the excess heat capacity were analyzed using the following simple model. The model assumes that $C_p$ of the film at temperatures above 160 K is a linear combination of the part of film which underwent glass-softening and the part of the film which did not undergo glass-softening. In other words,

$$C_p^{model} = M_{LF} \cdot C_p^{water*} + M_{SF} \cdot C_p^{ice*}, \qquad (10),$$

where $M_{LF}$ and $M_{SF}$ represent the masses of liquid and solid fraction of the film above 160 K.

Expressing the mass of the solid fraction through the masses of the liquid fraction and the mass of the film, equation (10) may be rewritten as

$$C_p^{model} = M_{LF} \cdot C_p^{water*} + (M_{film} - M_{LF}) \cdot C_p^{ice*}, \qquad (11),$$

where $M_{film}$ is the mass of an ASW aggregate.

In order to compare our model to the experimentally observed *excess* heat capacity, $\Delta C_p^{film}$, the heat capacity of ice was subtracted from the $C_p^{model}$, resulting in

$$\Delta C_p^{model} = M_{LF} \left( C_p^{water*} - C_p^{ice*} \right), \qquad (12)$$

The expression is further modified by introducing mass fraction of liquid part of the film, $\chi_{Lf}$,

$$\Delta C_p^{model} = \chi_{Lf} \cdot M_{film} \cdot \left( C_p^{water*} - C_p^{ice*} \right), \qquad (13),$$

where $\chi_{Lf}$ is the ratio of the mass of the liquid fraction to the overall mass of a ASW aggregates.

Figure 7 shows the *excess heat capacities* of ASW nano-aggregates of thicknesses ranging from 2 nm to 20 nm. The $\Delta C_p^{film}$ values were fit with equation (13) in the temperature range from 160 to 200 K. From the fitting, we obtained the $\chi_{Lf}$ values for the films of various thicknesses. The values of the liquid mass fraction are shown next to the fit curves in the figure. According to the analysis, $\chi_{Lf}$, decreases with the increase in the thickness of the film. This behavior is inconsistent with glass softening taking place in the bulk, because for a bulk process $\chi_{Lf}$ would remain constant irrespective of the mass of the film. The alternative hypothesis is that the glass softening occurs at the surface of the film, which corroborates with the analysis.

Assuming that the glass-softening is confined to a thin layer on the free surface of the ASW aggregates, we can confirm the thickness of the thin layer. The mass fraction of liquid, $\chi_{Lf}$, is the ratio of the mass of the surface layer, $M_{surface\ layer}$ to the mass of the entire film, $M_{film}$.

$$\chi_{Lf} = \frac{M_{surface\ layer}}{M_{film}}, \qquad (14)$$

The mass of the surface layer can be expressed as,

$$M_{surface\ layer} = Th_{LL} \cdot \rho_{H_2O} \cdot S_{area}, \qquad (15)$$

where $Th_{LL}$ represents the thickness of the surface layer, $\rho_{H_2O}$ is the density of water and $S_{area}$ surface area of the layer.

As described earlier, the surface area, $S_{area}$, of the nano-aggregates of ASW relative to the geometrical surface area of the filament can be determined, by varying the magnitude of the fitting parameter α for films ranging in thicknesses from 2 nm to 20 nm.

Combining equation (14) & (15), we obtain

$$\chi_{Lf} = \frac{Th_{LL} \cdot \rho_{H_2O} \cdot S_{area}}{M_{film}}, \qquad (16)$$

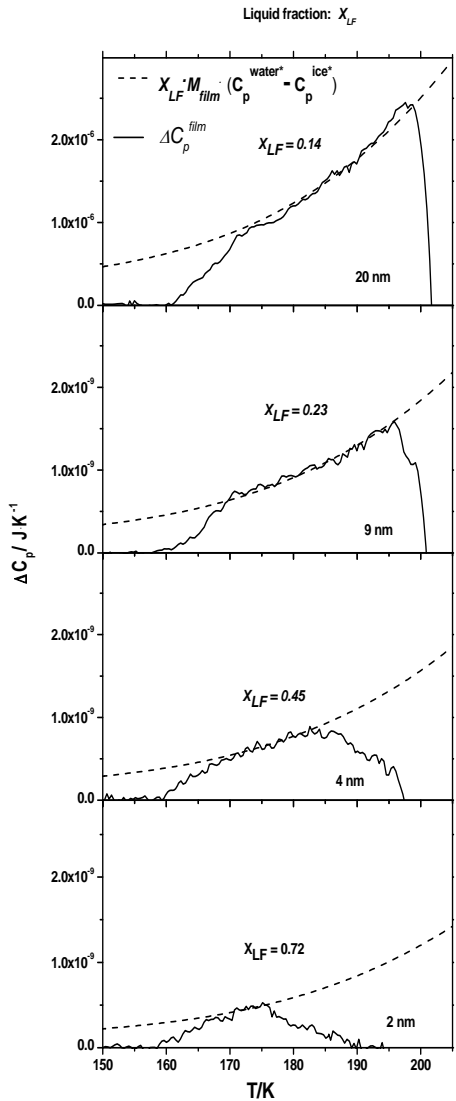

**Figure. 7.** Excess heat capacities (blue line) of ASW nano-aggregates of thicknesses ranging from 2 nm to 20 nm. Each thermogram is fitted with a model, equation (13). See text for details.

Figure 8 shows a plot of $\chi_{Lf}$ as a function of $\frac{\rho_{H_2O} \cdot S_{area}}{M_{film}}$. The plot is linear. The slope of the plot gives the thickness of the liquid layer after glass-softening, which is approximately 4-5 ML. *Therefore, we conclude that during an FSC experiment, the glass-softening of nanoaggregates of ASW occurs in the surface layer.*

We have attributed the glass-softening of ASW nano-aggregates occurring at 160 K to the surface layer of the film. Nevertheless, the interface between ASW nano-aggregates and tungsten filament presents a locality where glass softening of ASW nano-aggregates may also occur. In our past studies[36] we observed glass-softening taking place at interfaces between benzene and ASW, and methanoic acid and ASW. In these experiments, the glass-softening was manifested as an endotherm with the onset near 180 K. In the present study, no such endotherm is observed at 180 K at the interface between tungsten and ASW nano-aggregates. The failure to observe glass endotherm at 180 K may be explained in four different ways.

*First*, unlike "soft" interface between benzene and ASW, the "hard" interface between tungsten filament and ASW aggregates may impede water's molecular kinetics.[32] The enthalpy relaxation time at a particular temperature should increase due to the impeding of ASW aggregates molecular kinetics. Therefore, the glass softening may take place at temperatures higher than 200 K. The possible glass softening taking place at temperatures above 200 K is not observable due to simultaneous crystallization of ASW taking place at 200 K. Thus, the endotherm due to glass-softening is not observed at 180 K.

*Second*, the tungsten filament may be covered by a thin layer of graphite, making its surface highly hydrophobic. Thus, the contact area between ASW nano-aggregates and the filament surface may be small due to large contact angle. This should result in few molecules of ASW undergoing glass-softening. Therefore, the resulting glass endotherm may be too small to detect.

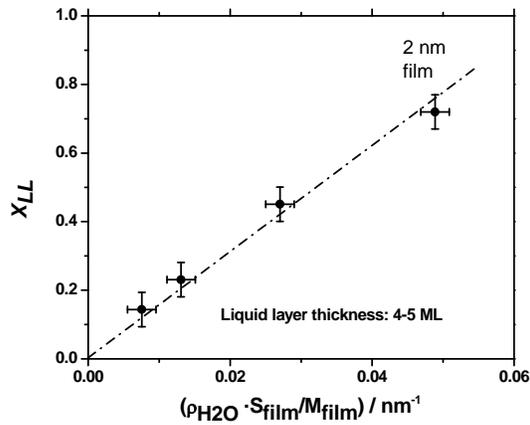

**Figure. 8.** $\chi_{Lf}$ as a function of $\frac{\rho_{H_2O} \cdot S_{area}}{M_{film}}$. The slope of the plot gives the thickness of the liquid layer after glass-softening, which is approximately 4-5 ML. See text for details.

*Third*, even in the case of 20 nm thick ASW films, the nano-aggregates may not be completely annealed. The complete annealing of the ASW would have resulted in sintering of the nano-aggregates. Because sintering, ripening or pore collapse of ASW nanoaggregates are exothermic processes, the endothermic process due to glass-softening may be masked.

*Finally*, surface tension may play a more important role in ASW nano-aggregates compared to those in thin films of ASW,[36]. The high surface curvature of ASW nano-aggregates must result in the decrease in their surface tension. The resulting decrease in the surface tension of ASW nano-aggregates must ultimately result in a higher molecular mobility. From the standpoint of heat capacity spectroscopy (FSC or DSC), higher molecular mobility in ASW nano-aggregates must be manifested as a significant decrease in the enthalpy relaxation time at a particular temperature. Therefore, in ASW nano-aggregates, the glass transition must be observed at a temperature lower than 180 K. Therefore, it is possible that the glass-softening of ASW clusters may occur at 160 K.

**F. Surface glass transition of ASW.**

In our previous publications[36-38] we observed that bulk-like ASW is unable to undergo glass-softening prior to crystallization. Nevertheless, when confinement or interfaces was introduced in the bulk-like ASW samples, the resulting samples could undergo glass-softening at temperatures 20 K below onset of crystallization. In the first study[37], bulk-like water was doped with molecules like acetic acid, carbon tetrachloride, pentanol or ethanol. In the second study [36], multiple nano-scale films of water (50 nm thickness) were alternated with either benzene or methanoic acid films respectively. For both the studies we observed that ASW samples could undergo glass-softening at temperatures of about 20 K prior to onset of crystallization.

We rationalized the results in the following way. In bulk-like ASW samples, each water molecule is able to form four cohesive hydrogen bonds with neighboring water molecules. The molecular structure of well annealed bulk-like ASW is similar to that of crystalline ice. When an impurity molecule is added or an interface is introduced in pure bulk-like ASW, it disrupts the cohesive hydrogen-bond network present in ASW i.e., the water molecules are unable to form four hydrogen bonds. Due to the disruption of the hydrogen-bond network, the molecular mobility increases at any given temperature. This increase in molecular mobility is manifested as a significant decrease in enthalpy relaxation time at a given temperature. Therefore, in doped bulk-like ASW sample or interfaces introduced in bulk-like sample, glass endotherm is observable at temperatures 20 K below crystallization. The lack of detectable glass endotherms, in the case of bulk-like, glassy water, due to cohesive H-bonding in sample leading to increase in enthalpy relaxation time , is perfectly consistent with this hypothesis, because such transition must occur at temperatures tens of degrees higher, and therefore is masked by the onset of ASW crystallization. In summary, *surface glass transition* before onset of crystallization is observed for ASW samples in confined geometry whereas the same is not observed for pure bulk-like ASW samples

The results of our previous studies show that the introduction of interfaces or confinement in ASW leads to observation of glass-softening at temperatures of 20 K below onset of crystallization. The glass softening for 50 nm nano-scales of ASW or doped ASW (doped with ethanol, pentanol, carbon tetra-chloride or acetic acid) was observed at 180 K. On the other hand, for our current study, the glass-softening of 2 nm nano-aggregates of ASW was observed at 160 K. The 20 K more lowering of glass transition temperature in case of 2 nm nano-aggregates compared to that of 50 nm thick nano-scale ASW may be rationalized due to severe confinement effects of 2 nm aggregates of water.

The introduction of confinements or interfaces leading to *surface glass transition* phenomena is well recognized for a wide variety of materials[44]. For example, the existence of surface or interfacial glass transitions in thin films of polymers is already well recognized in polymer science community[45]. Furthermore, the values of surface glass transition temperature tend to be tens of degrees lower than bulk (standard) glass transition temperature. For example, using first principle and classical MD studies for ASW films of thicknesses 4 nm, it has been observed that glass transition in these samples take place at 30 K corresponding to the bulk samples, which corroborates with our present results[46]. Lastly, the results of Devlin and coworkers show that the surface of nanoscale ice particles acts as source of Bjerrum defects[47], thus increasing the rate of structural relaxation in the 2-3 nm "subsurface" region even at cryogenic temperatures[12, 48]. These results provide significant support to surface glass transition phenomena of nano-aggregates of ASW of our present study.

Sepulveda and co-workers investigated glass transition phenomena in ultrathin films of ASW, size ranging from 16- 150 nm[49]. At a heating rate of $2.4 \cdot 10^4$ K/s, they observed glass transition in ASW to occur at 174 K. Our results show that at a heating rate of $1.1 \cdot 10^5$ K/s, the ASW nano-aggregates underwent softening at 160 K and had a vaporization endotherm at 185 K. We would like to point out that the 14 K difference in onset of Tg may have risen due to taking into different accounts of sublimation rate of water. It is well established that below 160 K, the sublimation rate of ASW is twice that of crystalline ice[43]. In our analysis, we took into consideration both the sublimation rate of ASW and crystalline ice. Sepulveda et. al. [49] took into consideration of sublimation rate of crystalline ice only in their analysis. This may have led to differences in interpretation of our results.

IV SUMMARY AND CONCLUSIONS

We developed and used a simple model to disentangle the contribution of various endo- and exothermal processes to the complex thermograms of ASW nano-aggregates. This model takes into account not only the possible glass softening transitions but also vaporization of the ASW samples. Following results were obtained when calorimetry scans of ASW nano-aggregates were analyzed using this model. The endotherm at 160 K was assigned to glass-softening of ASW nano-aggregates. The endotherms at 185 and 220 K were assigned to vaporization of ASW nano-aggregates. The exotherm at 175 K may arise due to Ostwald ripening of ASW nano-aggregates. Finally, the exotherm at 200 K is due to crystallization of ASW nano-aggregates. We also find that glass-softening for nano-aggregates starts at their surface.

We demonstrate unambiguously that nano-aggregates of glassy water are distinct from the bulk-like samples in terms of glass-softening dynamics: while nano-aggregates of water undergoes a glass transition at temperatures 40 K below the onset of crystallization, no such transition exist in the case of bulk glassy water. We attribute these differences to likely surface, which is apparent only in nano-aggregates of glassy water characterized by high surface to volume ratio.

Taking into account highly plausible interfacial glass transition of water and the fact that it may occur at temperatures tens of degrees lower than the bulk glass transition, the data from studies of water dynamics conducted with ultrathin films or using surface sensitive techniques must be interpreted with caution. In particular, results on molecular dynamics in deeply supercooled water should be examined for a possible impact of interfacial phenomena in the discourse of fundamental properties of bulk-like, defect-free, and pure (i.e., "ideal") condensed phases of $H_2O$. Because preparation of such ideal glassy samples is challenging, possible contribution of interfacial processes to observed phenomena should be considered not only in the case of experiments with ultrathin films, but also in the cases of certain bulk-like samples.

Finally, we emphasize that despite significance of research into "ideal" condensed phases of $H_2O$, studies of glass transition dynamics in interfacial water and aqueous solutions are equally important due to the prevalence of these in natural environments and in biological systems. Therefore, the interfacial or surface glass transition phenomena in water and other condensed phase systems is an exciting area of experimental and theoretical research. It is the authors' belief that eventual resolution of disagreement on glass transition dynamics of water would not devalue a particular group of contrasting results but would rather make it possible to better define their relevance and applicability to a particular class of natural phenomena.

*Acknowledgments:* This work was supported by National Science Foundation grant 1012692